\documentclass[onecolumn,  numbers]{els-mrw} 

\usepackage{amsmath,amssymb,amsfonts,amsthm,makeidx,graphicx}
\usepackage{txfonts}
\usepackage{helvet}
\usepackage{hyperref}
\usepackage{slashed}

\begin{document}


\chapter{Introduction to Lattice Field Theory  }\label{chap1}

\author[1,2]{ Ra\'ul A. Brice\~no}
\address[1]{Department of Physics, University of California, Berkeley, CA 94720, USA} 
\address[2]{Nuclear Science Division, Lawrence Berkeley National Laboratory, Berkeley, CA 94720, USA}

\articletag{Chapter Article tagline: update of previous edition, reprint.}

\maketitle

\begin{abstract}[Abstract]
This chapter provides a pedagogical introduction to lattice quantum field theory, with strong emphasis on lattice quantum chromodynamics. The chapter reviews key foundational concepts of lattice quantum chromodynamics, as well as a broad summary of ongoing research in the field. 
\end{abstract}

 \begin{keywords}
 	  lattice QCD, hadron structure, hadron spectroscopy, nuclear physics, finite temperature, g-2, inclusive rates, real-time methods 
\end{keywords}

\section*{Objective}
The objective is to give a brief introduction to: 
\begin{itemize}
	\item  lattice quantum chromodynamics,
        \item lattice as a regulator, 
        \item Wick rotation and discretization of the action, 
        \item definition and construction of correlation functions, 
        \item review of some applications of lattice quantum chromodynamics. 
	\end{itemize}

\section{Introduction \label{Sec}}

This chapter introduces some basic ideas of lattice field theory. Lattice field theories play a key role in a range of subfields in Physics. Within high-energy physics, the biggest motivation is its application to understand the consequences of the fundamental theory of the strong nuclear force, quantum chromodynamics (QCD), and to a lesser extent, to study possible extensions of the Standard Model of Particle physics that are non-perturbative. This chapter provides an introduction to the former of these two.

Theoretical efforts in understanding the phenomenological consequences of QCD are driven to some extent by two broad goals:
\begin{itemize}
\item understanding the emergence of the building blocks of matter from the fundamental interactions,
\item understanding the QCD contribution to experimental observables that could provide signals for new physics.
\end{itemize}
Each is interesting in its own right. Before discussing how lattice QCD can play a role in both of these lines of research, we elaborate on why these remain active lines of research.  

For those not so familiar with the theory, it might be surprising to know that to this day, there is no systematically improvable analytic procedure for accessing the spectrum of QCD. This is perhaps most striking when considered against the backdrop of two facts. First, the vast majority of particles observed in particle accelerators for the last century are hadrons, i.e., bound states of quarks and gluons~\cite{ParticleDataGroup:2024}. Of these, only a small fraction of their masses have been reproduced directly from QCD. Second, it has now been over 50 years since a series of seminal theoretical works~\cite {Fritzsch:1973pi, Gross:1973id, Politzer:1973fx} resulted in the conclusion that QCD must describe hadrons at the fundamental level. The primary obstacle to accessing and ultimately understanding the spectrum of QCD is its non-perturbative nature at low to moderate energies. The fact that the theory is non-perturbative not only precludes the determination of the spectrum, but it also puts tight restrictions on the class of observables that can be determined directly. This includes an endless number of observables that have been obtained in terrestrial experiments, but also quantities that would give insight into the evolution and properties of astrophysical systems.   

The second broad stroke motivation for constraining QCD observables is to remove the dominant Standard Model ``background" of a quantity that may provide signals for potential new physics. Two notable examples of these are precision electroweak matrix elements and the anomalous magnetic moment of the muon, both of which are discussed further below. 

The advent of lattice QCD is now making a large number of previously inaccessible observables numerically determinable from the theory. Traditional use of lattice QCD using classical computing is generally defined in a finite, discretized Euclidean spacetime. This allows for some quantities to be determined using statistically improvable approximations, which, in general, make soft assumptions about the strength of the coupling. This chapter reviews some of the key concepts behind the foundations of lattice QCD and lattice field theories in general. It discusses some observables that can be immediately accessed and explains why, and finally, it discusses exciting lines of research that are expanding this ever-growing field. This is by no means meant to be a complete account of this rather rich field. For readers seeking a more detailed introduction to lattice field theory and its application to QCD, several comprehensive textbooks are available~\cite{Rothe:2012, Gattringer:2010, DeGrand:2006,  Creutz:1983}. These provide in-depth discussions of both the formalism and practical aspects of lattice simulations.

\section{Lattice: The most natural regulator
\label{Sec
}}

At this point, it is hopefully clear to the reader that in studying quantum field theories (QFTs), one will generally encounter Ultraviolet (UV) singularities. These UV singularities arise when integrating over four-momenta that go to values much higher than we expect our QFT to hold. As a result, these singularities cannot be physical, and they require careful treatment. Intuitively, the most natural tool to regulate these singularities is to introduce a UV cutoff that removes high-energy modes.

Motivated to develop a framework to understand the phenomena of confinement in QCD, Kenneth Wilson~\cite{Wilson:1974sk} adopted insights from condensed matter systems to regulate QFTs by introducing a lattice spacing ($a$). In condensed matter systems, the lattice spacing is a scale that has physical meaning, e.g., the distance between adjacent atoms. Wilson's insight was to introduce an artificial spacing in both space and time that would serve as a UV regulator. This regulator has additional advantages. First, it is still the only regulator for which one can define the measure in the path-integral representation of any QFT. Second, to date, it has provided the only systematically improvable method for numerically evaluating QCD observables.

In practice, to study numerical consequences of the theory using lattice QCD, one requires the introduction of not just a UV regulator, but also an Infrared (IR) regulator, namely the spacetime volume ($V$). This is done so that the mathematical objects appearing in the theory can be represented as finite-sized tensors that can be manipulated numerically.

QCD is perfect for this setup for two key reasons. First, the theory is asymptotically free in the UV region. This means that the QCD coupling goes to zero for large energy scales, or equivalently, small spacetime separations. This implies that effects associated with the choice of the UV cutoff can be dealt with perturbatively for sufficiently small values of $a$. Second, QCD is confining and has a mass gap. This means that the IR degrees of freedom are not quarks and gluons, but rather hadrons. Furthermore, the lightest hadron (the pion) has a non-zero mass ($m_\pi$), at least away from the chiral limit. This means that the range of interactions of the theory scales as $\mathcal{O}(1/m_\pi)$, and as a result, one expects that one may be able to directly constrain a large class of quantities using finite volumes. To make a more specific statement, it is useful to consider a specific $4D$ lattice with a volume $V=T\times L^3$. Namely, let the temporal extent be $T$, and the spatial component be cubic with a length $L$. The statements above can be translated to mean that there is a natural expectation that many observables will converge quickly to their physical values as one continuously increases $m_\pi L$ and $m_\pi T$. The exponential convergence of the spectrum of stable particles with the IR regulator was first proved by L\"uscher~\cite{Luscher:1985dn}. Not all observables converge exponentially fast. Most recently, Ref.~\cite{Burbano:2025pef} provided a derivation of the volume scaling for any observable.

In Sec.~\ref{sec:action}, we will discuss the natural procedure for discretizing QCD while preserving gauge invariance. The basic idea is to introduce ``gauge links" that are elements of the SU(3) gauge group of QCD. This is just a gauge transporter that connects nearby lattice sites, while the Fermions reside in the lattice sites. Before discussing details of the discretization, let us first discuss one key ingredient of numerical evaluations, namely the need for making the spacetime Euclidean.

\section{Wick rotating to Euclidean spacetime}
It is common to perform a Wick rotation to facilitate numerical simulations, transforming the theory from Minkowski spacetime to Euclidean spacetime. This rotation can be performed explicitly by taking time $t$ and replacing it with $-i \tau$, where $\tau$ is real. Under this prescription, the Minkowski partition function, $Z_M$, is transformed to the Euclidean one, $Z_E$,  
\begin{equation}
Z_M = \int \mathcal{D}U \, \mathcal{D}\bar{\psi} \, \mathcal{D}\psi  e^{iS_M[U, \bar{\psi}, \psi]}\Rightarrow Z_E= \int \mathcal{D}U \, \mathcal{D}\bar{\psi} \, \mathcal{D}\psi  e^{-S_E[U, \bar{\psi}, \psi]},
\end{equation}
where $U$ denotes the gauge link variables on the lattice, $\psi$ and $\bar{\psi}$ are the quark fields, and $S_E$ is the Euclidean action of QCD.

This rotation simplifies the treatment of the path integral by converting oscillatory integrals into exponentially damped ones, making them more amenable to numerical evaluation. Formally, this replaces the $4D$ quantum field theory with a $3D$ quantum statistical field theory with a temperature $(T_{\rm temp.})$ fixed by the inverse of the temporal extent of the lattice $T$ ($T_{\rm temp.} \equiv (Tk_B)^{-1}$, where $k_B$ is the Boltzmann constant). 

In practice, for any  Euclidean spacetime, the path integral becomes a statistical field theory problem, where the weight of each configuration is $e^{-S_E[U, \bar{\psi}, \psi]}$. As we review in Sec.~\ref{sec:fermionic}, the fermionic part of the QCD action is quadratic in the fields. This means that one can integrate out the fermions exactly. Doing so, the Euclidean partition function becomes
\begin{equation}
Z_E = \int \mathcal{D}U \, \det M[U] \, e^{-S_g[U]},
\end{equation}
where $S_g[U]$ is the gauge field action and $M[U]$ is the Dirac operator matrix for the quarks in the background of the gauge fields~\cite{ DeGrand:2006}. The expectation value of an observable $\mathcal{O}$ can then be expressed as
\begin{equation}
\langle \mathcal{O} \rangle_E = \frac{1}{Z_E} \int \mathcal{D}U \, \mathcal{O}[U] \, \det M[U] \, e^{-S_g[U]}. 
\end{equation}

It is not known how to evaluate the remaining integral analytically. Given the high dimensionality of the integral, the only systematic approach for estimating such an integral is Monte Carlo sampling, where the probability density is taken to be proportional to $\det M[U] \, e^{-S_g[U]}$. For this to be possible, one must require that the action considered for the fermions is such that $\det M[U]>0$. Assuring this, one can then evaluate $N$ representative field configurations distributed according to $\det M[U] \, e^{-S_g[U]}$. At this point, one can then approximate the expectation value of the operator $\mathcal{O}$ by, 
\begin{align}
\langle \mathcal{O} \rangle_E \approx 
\frac{1}{N} \sum_n^N \mathcal{O}[U_n].
\end{align}

Key references in the development and application of Monte Carlo methods in Lattice QCD include the pioneering work by Creutz (1980)\cite{Creutz:1980zw} and the comprehensive review by DeGrand and DeTar (2006)\cite{DeGrand:2006}, which provide foundational insights and methodologies for these simulations.
The implementation of Monte Carlo sampling in Lattice QCD typically involves algorithms such as the Metropolis-Hastings algorithm, Hybrid Monte Carlo (HMC)~\cite{Duane:1987de}, and the more recent advancements like the Rational Hybrid Monte Carlo (RHMC) algorithm. These algorithms are designed to generate a sequence of lattice field configurations that are distributed according to the QCD action, ensuring that the statistical ensemble accurately represents the quantum field theory. The efficiency and accuracy of these algorithms are paramount, as they directly impact the precision of the computed physical observables such as hadron masses, decay constants, and the QCD phase diagram. Advances in computational power and algorithmic techniques continue to push the boundaries of what can be achieved with Lattice QCD, making Monte Carlo sampling an indispensable tool in the quest to understand the fundamental forces of nature.

By having Wick rotated the theory, one may na\"ively conclude that physically interesting quantities are out of reach. In Sec.~\ref{sec:application}, we explain how, despite being in a Euclidean spacetime, one can still obtain a large class of physical quantities, including ones that may be time-sensitive. 

\subsection{Symmetry breaking}
By introducing two regulators and Wick rotating the theory, we have broken several symmetries of the original theory. This can have profound consequences at the stage of identifying the correspondence between the observables that are directly accessible via lattice QCD and the desired physical quantities. For some quantities, e.g. masses of stable particles, this introduces systematic errors that can be corrected by taking the physical point limit (defined in Sec.~\ref{sec:2ptCorr}). For other quantities, e.g. parton distribution functions, these lattice artifacts make direct determination impossible. For such cases, it may be possible to find indirect correspondences that bypass such limitations. In the example of parton distribution functions, see Refs.~\cite{Detmold:2005gg,Ji:2013dva,Radyushkin:2016hsy}.

\section{Discretization of the Action}
\label{sec:action}
In lattice field theories, the discretization of the action is a crucial step that allows for the numerical evaluation of the theory. The QCD action consists of both fermionic and gluonic components, each of which must be carefully discretized to maintain the properties of the continuum theory while ensuring numerical stability and accuracy. In general, one can construct an infinite number of actions on the lattice that recover the standard QCD action in the continuum. Here we discuss the simplest of these, first introduced by Wilson. We begin by first introducing the gauge link. 

\subsection{Gauge links}
In lattice QCD, the gauge fields are discretized and represented by gauge links (also called link variables), which serve as the fundamental degrees of freedom for the gluon fields on the lattice. Instead of working with the continuous gauge field $A_\mu(x)$, the lattice formulation assigns to each oriented link between neighboring lattice sites a group element $U_\mu(x)$, which is an element of the gauge group (for QCD, $SU(3)$). Specifically, $U_\mu(x)$ is associated with the link starting at site $x$ and pointing in the $\mu$ direction.
The gauge link is defined as the path-ordered exponential of the gauge field along the link. If we require the path to be a straight line connecting two adjacent sites, one at $x$ and the other at $x+a\hat{\mu}$, we can write the link explicitly as,
\begin{equation}
U_\mu(x) =  \exp \left[ i g \int_{x}^{x+a\hat{\mu}} A_\mu(y) dy \right],
\end{equation}
where $g$ is the gauge coupling. For a sufficiently small value of $a$, we can approximate this integral as,
\begin{equation}
U_\mu(x) \approx \exp\left[ i g a A_\mu(x) \right].
\end{equation}
Gauge invariance is a fundamental principle in QCD, and the lattice formulation must respect this symmetry. Under a local gauge transformation $\Omega(x) \in SU(3)$, the gauge links transform as
\begin{equation}
U_\mu(x) \rightarrow U'_\mu(x) = \Omega(x) \,  U_\mu(x) \, \Omega^\dagger(x + a\hat{\mu}),
\end{equation}
where $\Omega(x)$ is the gauge transformation at site $x$, and $\Omega^\dagger(x + a\hat{\mu})$ is its Hermitian conjugate at the neighboring site in the $\mu$ direction. This transformation ensures that closed loops of gauge links, such as the Wilson loop or the plaquette, are gauge invariant.
For example, the smallest Wilson loop, known as the plaquette, is defined as
\begin{equation}
U_{\mu\nu}(x) = U_\mu(x)\,  U_\nu(x + a\hat{\mu})\,  U^\dagger_\mu(x + a\hat{\nu})\,  U^\dagger_\nu(x),
\end{equation}
which transforms under gauge transformations as
\begin{equation}
U_{\mu\nu}(x) \rightarrow \Omega(x) \, U_{\mu\nu}(x) \, \Omega^\dagger(x),
\end{equation}
so that the trace, $\mathrm{Tr}[U_{\mu\nu}(x)]$, is gauge invariant.
The gauge links thus play a central role in constructing lattice actions and observables that respect the local gauge symmetry of QCD.

\subsection{Fermionic Action}
\label{sec:fermionic}
The fermionic part of the QCD action describes the dynamics of quarks, which are represented by Dirac fields. The continuum fermionic action is given by
\begin{equation}
S_F = \int d^4x \, \bar{\psi}(x) (\slashed{D}  + m) \psi(x),
\end{equation}
where \( D \!\!\!\!/ = \gamma^\mu D_\mu \) is the Dirac operator in the chiral limit and \( D_\mu = \partial_\mu - i g A_\mu \) is the covariant derivative. Discretizing the fermionic action poses significant challenges due to the \emph{fermion doubling problem}, where naive discretization leads to additional, unphysical fermion species. One common solution is the Wilson fermion formulation, introduced by Kenneth Wilson~\cite{Wilson:1974sk}, which adds a term to the action that removes the unwanted doublers at the cost of explicitly breaking chiral symmetry:
\begin{equation}
S_F^{\text{Wilson}} = a^4 \sum_x \left[ \bar{\psi}(x) \left(m + \frac{r}{2a}\right) \psi(x) + \frac{1}{2a} \sum_\mu \left( \bar{\psi}(x) (\gamma_\mu - r) U_\mu(x) \psi(x + a\hat{\mu}) - \bar{\psi}(x + a\hat{\mu}) (\gamma_\mu + r) U_\mu^\dagger(x) \psi(x) \right) \right].
\end{equation}
Another approach is the staggered fermion formulation~\cite{Kogut:1974ag}, which reduces the number of doublers by distributing the components of the Dirac spinor over different lattice sites, preserving a remnant of chiral symmetry. More sophisticated formulations, such as domain wall fermions and overlap fermions, have been developed to preserve chiral symmetry more accurately, at the expense of increased computational complexity~\cite{Kaplan:1992bt, Narayanan:1993sk}.

\subsection{Gluonic Action}

The gluonic part of the QCD action describes the dynamics of the gauge fields, which mediate the interactions between quarks. The standard discretization of the gluonic action is the Wilson plaquette action, which approximates the field strength tensor using the smallest closed loop on the lattice, known as the plaquette:
\begin{equation}
S_G = \frac{2}{g^2} \sum_{x, \mu < \nu} \left[ 1 - \frac{1}{N_c} \text{Re} \, \text{Tr} \, U_{\mu\nu}(x) \right],
\end{equation}
where \( U_{\mu\nu}(x) \) is the plaquette operator defined above.
This formulation respects gauge invariance and is relatively simple to implement. More accurate discretizations, such as the Symanzik-improved action and the Iwasaki action, include larger loops to reduce discretization errors and improve the approach to the continuum limit~\cite{Symanzik:1983dc, Iwasaki:1983iya}.

\subsection{Chiral Symmetry and Discretization}

The interplay between chiral symmetry and the discretization of the fermionic action is a central concern in lattice QCD. Chiral symmetry is an important feature of the QCD Lagrangian in the massless limit, and its breaking can lead to significant artifacts in the simulations. The Nielsen-Ninomiya theorem states that any local, translationally invariant lattice formulation with exact chiral symmetry and no fermion doubling is impossible~\cite{Nielsen:1981hk}. This theorem implies that preserving chiral symmetry on the lattice requires either accepting fermion doubling or breaking chiral symmetry explicitly.

Formulations such as domain wall fermions and overlap fermions address this issue by introducing an additional dimension or using a modified Dirac operator that satisfies the Ginsparg-Wilson relation, which ensures an exact lattice chiral symmetry~\cite{Ginsparg:1981bj}. These approaches allow for the restoration of chiral symmetry in the continuum limit and provide a more faithful representation of QCD.

In summary, the discretization of the QCD action on the lattice involves careful consideration of both the fermionic and gluonic components. Various formulations and improvements have been developed to address the challenges of fermion doubling, chiral symmetry breaking, and numerical stability, making lattice QCD a powerful and precise tool for studying the strong interaction.

\section{Correlation functions and propagators}
\label{sec:application}
Lattice QCD has proven to be a remarkably powerful tool to study the consequences of QCD, as well as other parts of the Standard Model. This chapter cannot do justice to the vast literature of this field. Instead, it provides some conceptually simple examples illustrating its reliability and versatility. We will pay close attention to two-point correlation functions, since these provide the most useful mechanism for constraining the spectrum of the theory. 

\subsection{Two-point correlators and spectroscopy}
\label{sec:2ptCorr}
If we first ignore the fact that, in general, the temporal extent of a lattice will need to be finite, we can define two-point correlation functions as the expectation value of two operators displaced in  time:
\begin{equation}
C_{\rm 2pt.}(t) \equiv \langle \mathcal{O}(t)\, \mathcal{O}^\dag(0) \rangle_E ,
\label{eq:C2pt}
\end{equation}
where $\mathcal{O}(t)$ is a generic operator with the quantum numbers of a state we wish to study. For theories like QCD, we do not know the exact form that such operators should take to create a given state. Instead, we can construct so-called ``\emph{interpolating operators}", which will couple to all states with the same quantum numbers. Within QCD, these operators must be color neutral, i.e. gauge covariant. It is easy to imagine constructing a large set of operators that satisfy this criterion while being constructed from quark fields, covariant derivatives, and gauge links. 

In general, the two operators being considered do not need to be the same; in fact, there are many advantages in considering a large range of operators that are distinct and solving the so-called Generalized Eigenvalue Problem (GEVP)~\cite{Michael:1985ne,Luscher:1990ck}, which is the state of the art is modern-day analysis. The one requirement for such a correlation function to be non-zero is for the operator at the origin (sometimes called ``\emph{source}") and the operator at a later time $t$ (sometimes called ``\emph{sink}") to have the same quantum numbers. 

Identifying the operator $\mathcal{O}(t)$ as a Heisenberg operator in Euclidean time, allows one to write it as $e^{\hat{H}t}\mathcal{O}(0)\, e^{-\hat{H}t}$, where $\hat{H}$ is the Hamiltonian of the theory. Because the theory is formulated in a finite volume, the Hamiltonian is strictly speaking the finite-volume Hamiltonian, which has a discrete spectrum.  If we denote the eigenstates and eigenvalues of $\hat{H}$ as $|E_n\rangle $ and $E_n$, respectively, one finds that Eq.~\eqref{eq:C2pt} can be written as
\begin{equation}
C_{\rm 2pt.}(t) =\sum_n |\langle0| \mathcal{O}(0)|E_n\rangle |^2\, e^{-E_nt} , 
\label{eq:C2pt_v1}
\end{equation}
where $|0\rangle $ is the interacting vacuum of the theory, and the finite-volume states have been normalized to $1$. This is the so-called spectral decomposition of the correlation function. 

From this expression, one immediately sees that if one can evaluate such a correlation function for sufficiently large time extents, the correlator could be approximated to be dominated by the ground state with the quantum numbers of $\mathcal{O}(0)$. More specifically, for asymptotically large times, one finds, 
\begin{equation}
C_{\rm 2pt.}(t) \approx  |\langle0| \mathcal{O}(0)|E_0\rangle |^2\, e^{-E_0t}, 
\label{eq:C2pt_v2}
\end{equation}
where corrections scale as $e^{-(E_1-E_0)\,t}$. Making it clear that the correlator will be saturated by the ground state if $(E_1-E_0)\,t \gg 1$. 

One useful tool in the analysis of correlation functions is the so-called effective mass, $m_{\rm eff} (t)$ 
\begin{align}
a\,m_{\rm eff} (t) = \log\left(\frac{C_{\rm 2pt.}(t)}{C_{\rm 2pt.}(t+a)}\right), 
\end{align}
where the lattice spacing, $a$, ensures that all quantities are dimensionless. At large times satisfying $(E_1-E_0)\,t \gg 1$, the effective mass asymptotes to the energy of the ground state in lattice units, i.e. in units of $a^{-1}$. In the next section, we briefly discuss how one can determine the lattice spacing and thereby set the scale of a lattice calculation.

If the interpolating operator used has zero total momentum and the ground state is a stable particle, the energy can be interpreted as the mass of the particle at a non-zero value of the lattice spacing ($a$), a finite volume ($V$), and some choice of the values of the input parameters of the theory, which for QCD are just the quark masses $(\{m_q\})$, i.e. $E_0=m(a,V,\{m_q\})$. 

To then recover the physical mass of this particle, one would first need to make sure that $\{m_q\} $ are tuned to their physical values, $\{ m_q^{\rm phys.} \} $. Alternatively, one can repeat calculations at a range of values of $\{m_q\} $ to then extrapolate or interpolate observables to the physical point. Furthermore, one needs to repeat the calculation for a range of values of $a$ and $V$ to then extrapolate to the continuum ($a\to0$) and infinite-volume limit ($V\to \infty$),
\begin{align}
m^{\rm phys.}
=
\lim_{a\to 0}
\lim_{V\to \infty}
\lim_{\{m_q\}\to \{m_q^{\rm phys.}\}}
m(a,V,\{m_q\}).
\label{eq:physical_point}
\end{align}
This limit can be understood as the \emph{physical point limit}. Such extrapolations are generally guided by analytic expressions obtained using perturbative QCD or effective field theories (EFTs) of QCD, e.g. chiral perturbation theory. Such frameworks allow one to evaluate correlation functions using their \emph{diagrammatic representations}, describing systematic errors. For some discussion on such extrapolations and fits for precision measurements from lattice QCD, we point the reader to Flavour Lattice Averaging Group Review~\cite{Aoki:2024cqa}. 

In practice, the finite temporal extent requires imposing boundary conditions on the fields at the temporal boundaries. The fermionic fields are usually given anti-periodic boundary conditions, and the gauge fields are given periodic boundary conditions. This results in the two-point correlation functions taking on the form
\begin{equation}
C_{\rm 2pt.}(T,t) \equiv {\rm Tr} \left[e^{-\hat{H}T} \,\mathcal{O}(t)\, \mathcal{O}^\dag(0) \right]/{\rm Tr} \left[e^{-\hat{H}T}\right] , 
\end{equation}
where $T$ is the temporal extent of the lattice. From this expression, one sees that in the limit that $T$ goes to $\infty$, one recovers Eq.~\eqref{eq:C2pt}. For large but finite values of $T$, Eq.~\eqref{eq:C2pt} will receive corrections that must be taken into account in the analysis of the correlation function.

\subsection{Tuning quark masses and scale setting}

A crucial step in lattice QCD calculations is the tuning of quark masses and the determination of the lattice scale. Since simulations are performed in terms of bare parameters—such as the lattice gauge coupling and quark masses—these must be related to physical quantities to make contact with experimental results. The process generally involves adjusting the input parameters so that calculated observables, such as hadron masses, match their experimentally measured values.

Quark mass tuning is typically achieved by varying the bare quark mass parameters in the lattice action until the calculated masses of specific hadrons, usually the pion and kaon, agree with their physical values. In the isospin-symmetric limit, the up and down quark masses are set equal, and the pion mass is used as the primary tuning target. The strange quark mass is often tuned using the kaon or $\phi$ meson mass. For charm and heavier quarks, other hadron masses such as the $D$ or $J/\psi$ are employed. This tuning is performed iteratively, as the relationship between the bare quark masses and the resulting hadron masses is not linear and depends on the gauge coupling and lattice discretization effects.

It is important to note that there is nothing within QCD that gives us a sense of units. When evaluating consequences from the theory perturbatively, the dependence on the bare gauge coupling $g$ is replaced by the dependence on a dimensionful scale $\Lambda_{\rm QCD}$. This process of replacing a dimensionless quantity by a dimensionful one is known as \emph{dimensional transmutation}. By matching a theoretical prediction of, for example, a high-energy cross section to experimental data, one can constrain  $\Lambda_{\rm QCD}$. 

The concept of dimensional transmutation is also applicable to lattice QCD. There, the bare coupling of the action is more naturally replaced by the dimensionful lattice spacing $a$. Scale setting refers to determining the lattice spacing $a$, which sets the conversion between dimensionless lattice quantities and physical units. This is commonly accomplished by calculating a reference quantity whose value is well-known experimentally, such as the mass of the $\Omega$ baryon, the pion decay constant $f_\pi$, the Sommer parameter $r_0$ extracted from the static quark potential, and/or gradient flow techniques~\cite{Luscher:2010iy,BMW:2012hcm,MILC:2015tqx}. By matching the lattice result for this quantity to its physical value, the lattice spacing is determined. The choice of reference quantity can affect systematic uncertainties, so multiple methods are often employed for cross-checks.

As a concrete example, consider the $\Omega$ baryon mass. Using two-point correlators of the kind described in Eq.~\eqref{eq:C2pt}, one can constrain from the large time behavior $a m_\Omega^{\rm latt.}$, where we are emphasizing the fact that this value was obtained on the lattice. By requiring that this exactly equals the lattice spacing times the experimental value of the mass of the $\Omega$ baryon, $a m_\Omega^{\rm exp.}$, we obtain a definition of the lattice spacing,
\begin{align}
a \equiv  \frac{a m_\Omega^{\rm latt.}}{m_\Omega^{\rm exp.}}.
\end{align}
This might seem confusing at first, but it is important to emphasize that $a m_\Omega^{\rm latt.}$ is a dimensionless number that has been estimated numerically. As already emphasized, this is a choice. One could use other quantities to set the scale. Away from the physical point, defined by taking limits described in Eq.~\eqref{eq:physical_point}, different prescriptions for the lattice spacing can differ in the quoted quantities. At the physical point, all prescriptions must agree, unless there are additional underestimated systematic errors (e.g., fitting, excited state contamination).

\subsection{Wick contractions and propagators}

To compute the correlation functions, it is necessary to consider another representation. Having defined a correlation function as in, for example, Eq.~\eqref{eq:C2pt} in terms of quark and gauge fields, one can proceed to evaluate it using Wick's theorem. In particular, let 
\begin{equation}
C_{\rm 2pt.}(t) \equiv 
\langle \mathcal{O}(t)\, \mathcal{O}^\dag(0) \rangle_E  \approx
\frac{1}{N} \sum_{n=1}^N C_{\rm 2pt.}(t)[U_n],
\end{equation}
where $C_{\rm 2pt.}(t)[U_n]$ is the value of the two-point correlator for a fixed gauge field configuration $U_n$. Then, Wick's theorem tells us that $C_{\rm 2pt.}(t)[U_n]$ can be obtained by
\begin{align}
C_{\rm 2pt.}(t)[U_n]=
\text{sum over all Wick contractions for fixed $U_n$}.
\label{eq:C2pt_Wick}
\end{align}
In practice, this allows for the numerical determination of correlation functions. 

As an explicit example, let us consider arguably the simplest correlator, that of the  $\pi^+$ meson. A useful interpolating operator for the $\pi^+$ is $\pi^+(x) = \bar{d}(x)\gamma_5 u(x)$, where $u$ is the up-quark field and $d$ is the down-quark field. This operator has the correct isospin and parity. The two-point correlation function, which probes the propagation of the pion from a source to a sink separated in time, is then given by the expectation value $\langle \pi^+(x) \bar{\pi}^+(0) \rangle$ over the ensemble of gauge configurations. 

Applying Wick’s theorem, this four-fermion expectation value can be expressed in terms of quark propagators. Specifically, the contraction leads to a single connected diagram where the up and down quark fields are paired to form propagators. The result is a trace over spin and color indices of the product of the up and down quark propagators, sandwiched between $\gamma_5$ matrices: 
\begin{equation}
C_{{\rm 2pt.},\pi^+}(t)[U_n] = -\sum_{\mathbf{x}} \text{Tr}\left[M_u^{-1}(x,0)\,\gamma_5\, M_d^{-1}(0,x)\,\gamma_5\,\right][U_n].
\end{equation}
Here, $M_f^{-1}$ denotes the quark propagator for flavor $f$, and the sum over spatial positions projects onto definite momentum. An important point is that these propagators are generated in the fully interacting theory in the presence of the fixed gauge field $U_n$. 

\begin{figure*}[t]
    \centering
    \includegraphics[width=0.8\textwidth]{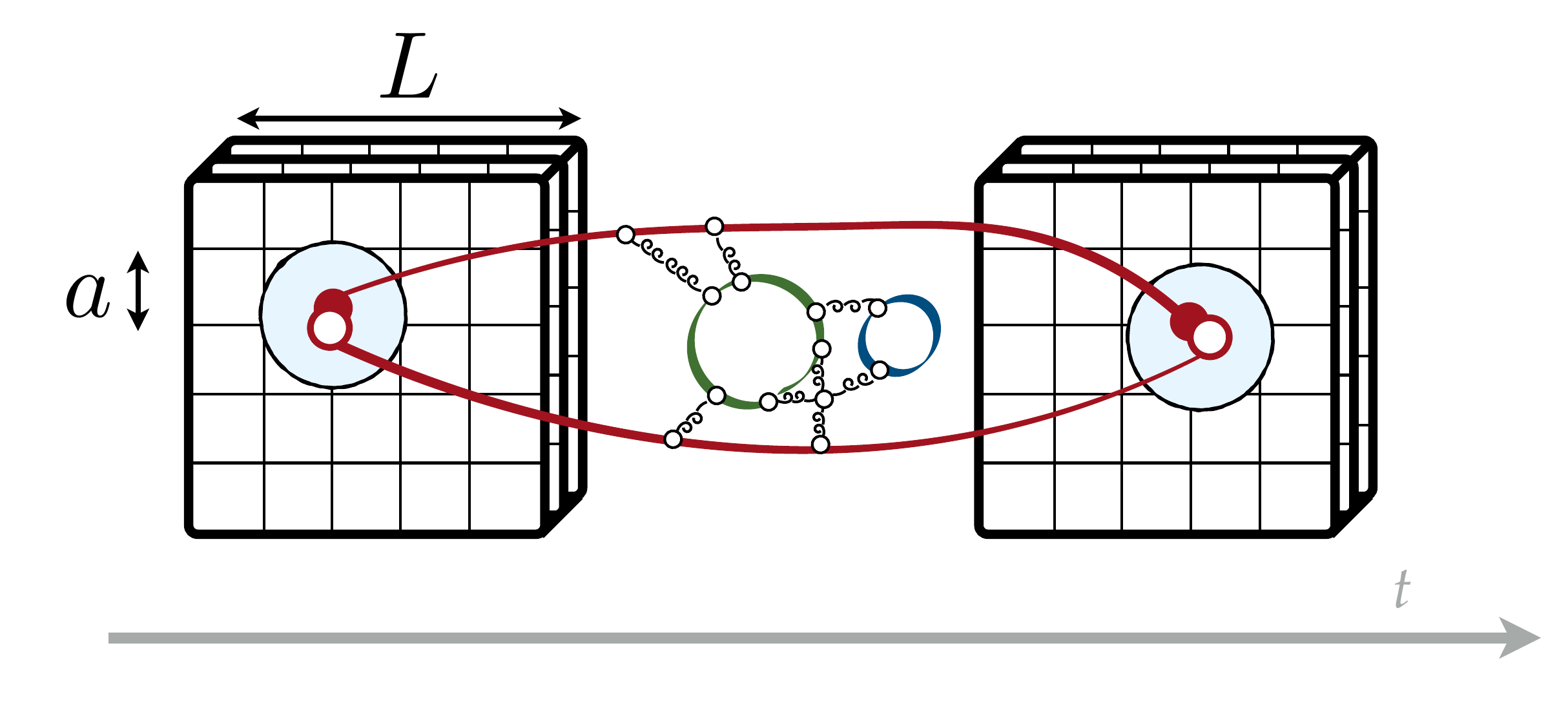}
    \caption{ Shown is a cartoon depiction of a possible contribution to the $\pi^+$ two-point correlator discussed in the main text.  
    }
    \label{fig:Cpi}
\end{figure*}

For a cartoon depiction of this correlation function, see Fig.~\ref{fig:Cpi}. The spatial extent of the lattice is shown as a cubic discrete volume. The quark and anti-quark fields are placed at the same lattice site, depicted with a solid and an open circle, respectively. The temporal extent is shown as a horizontal axis labeled $t$. The lines connecting quark fields at different times represent the propagators. To emphasize that these are propagators in the interacting theory, additional gluon field couplings to sea quarks are depicted.

A key computational simplification arises from $\gamma_5$-Hermiticity of the Dirac operator and the isospin symmetry of the light quarks, allowing the correlator to be evaluated using point-to-all propagators---propagators from a fixed source to any point on the lattice. This avoids the need to compute the full all-to-all propagator, significantly reducing the computational cost. As a result, the $\pi^+$ correlator is dominated by a single connected contraction, making it one of the simplest and most accessible quantities to calculate in lattice QCD studies.

\subsection{Propagators}

In lattice QCD, a point-to-all propagator refers to the quark propagator that originates from a fixed source point and extends to all other points on the lattice. To compute this object, one must solve the discretized Dirac equation for a given gauge field configuration. Specifically, for a quark of flavor $f$, the Dirac equation can be written as $M_f \psi = \eta$, where $M_f=(\slashed{D}  + m)_f$ is the Dirac operator (which encodes the lattice action and gauge links), $\eta$ is a source vector, and $\psi$ is the resulting quark field.

To construct a point-to-all propagator, the source $\eta$ is chosen to be nonzero only at a single lattice site (the ``point''), typically with unit entries in a selected spin and color component. The linear system $M_f \psi = \eta$ is then solved numerically, yielding $\psi$, which gives the propagator from the source point to every other point on the lattice for the chosen spin and color. This process is repeated for each spin and color component of the source, ultimately building the full set of propagators needed for contraction into hadronic correlation functions.

In practice, the inversion of the Dirac operator is performed using iterative algorithms such as the conjugate gradient method or its variants, which are well-suited for large, sparse systems \cite{DeGrand:2006, Gattringer:2010}. The efficiency of these solvers is often improved by techniques such as even-odd preconditioning, deflation, or multi-grid methods. Modern lattice QCD codes, such as \texttt{Chroma}~\cite{Edwards:2004sx} and \texttt{QUDA}~\cite{Clark:2009wm}, implement these algorithms on high-performance computing architectures, enabling the calculation of point-to-all propagators for realistic lattice sizes and quark masses. Despite these advances, the computation remains resource-intensive, and optimizing the inversion process is an active area of research.

\subsection{Three-point correlators and matrix elements}

Finally, we consider the second most widely studied class of correlation functions. Three-point correlation functions probe transition amplitudes between states, enabling the extraction of matrix elements of operators. Such matrix elements are key to constrain QCD contributions to electroweak decays and electromagnetic form factors, among many different quantities. 

An example of a three-point function is the following,
\begin{align}
C_{\rm 3pt.}(t, t_c) = \langle 0 | \mathcal{O}(t) \mathcal{J}(t_c) \mathcal{O}^\dagger(0) | 0 \rangle,
\end{align}
where $\mathcal{J}$ is the operator whose matrix elements we want to determine. In some special cases, this is a conserved current. 

The spectral decomposition of $C_{\rm 3pt.}$ can be obtained by identifying both $\mathcal{O}(t)$ and $\mathcal{J}(t_c)$ as Heisenberg operators in Euclidean time and then inserting two complete sets of states, 
as a sum over intermediate states:
\begin{align}
C_{\rm 3pt.}(t, t_c)= \sum_{n,m}\, \langle 0 | \mathcal{O}(0) | E_n \rangle \langle E_n | \mathcal{J}(0) | E_m \rangle \langle E_m | \mathcal{O}^\dagger (0)| 0 \rangle \,e^{-E_n (t-t_c)}\, e^{-E_m t_c}.
\end{align}
At large time separations, the ground-state matrix element $\langle 0 | \mathcal{O}(0) | E_0 \rangle \langle E_0 | \mathcal{J} (0)| E_0 \rangle \langle E_0 | \mathcal{O}^\dagger (0)| E_0 \rangle$ dominates, allowing for the extraction of physical observables such as form factors and decay constants. Careful analysis is required to isolate the desired contributions and control excited-state contamination.

\section{Modern-day research}

The field of lattice quantum field theory has matured significantly, enabling a wide variety of high-precision calculations that underpin our understanding of the Standard Model and guide searches for new physics. While it is impossible to review all developments here, we highlight some representative themes that illustrate the breadth and vitality of current research.

\textbf{Precision calculations:}  
A major success of lattice QCD has been the computation of fundamental quantities such as hadron masses, decay constants, and quark masses with percent-level or better uncertainties. These results provide stringent tests of QCD and are compiled in regular reviews such as those by the FLAG collaboration~\cite{Aoki:2024cqa}. 

\begin{figure*}[t]
    \centering
    \includegraphics[width=0.95\textwidth]{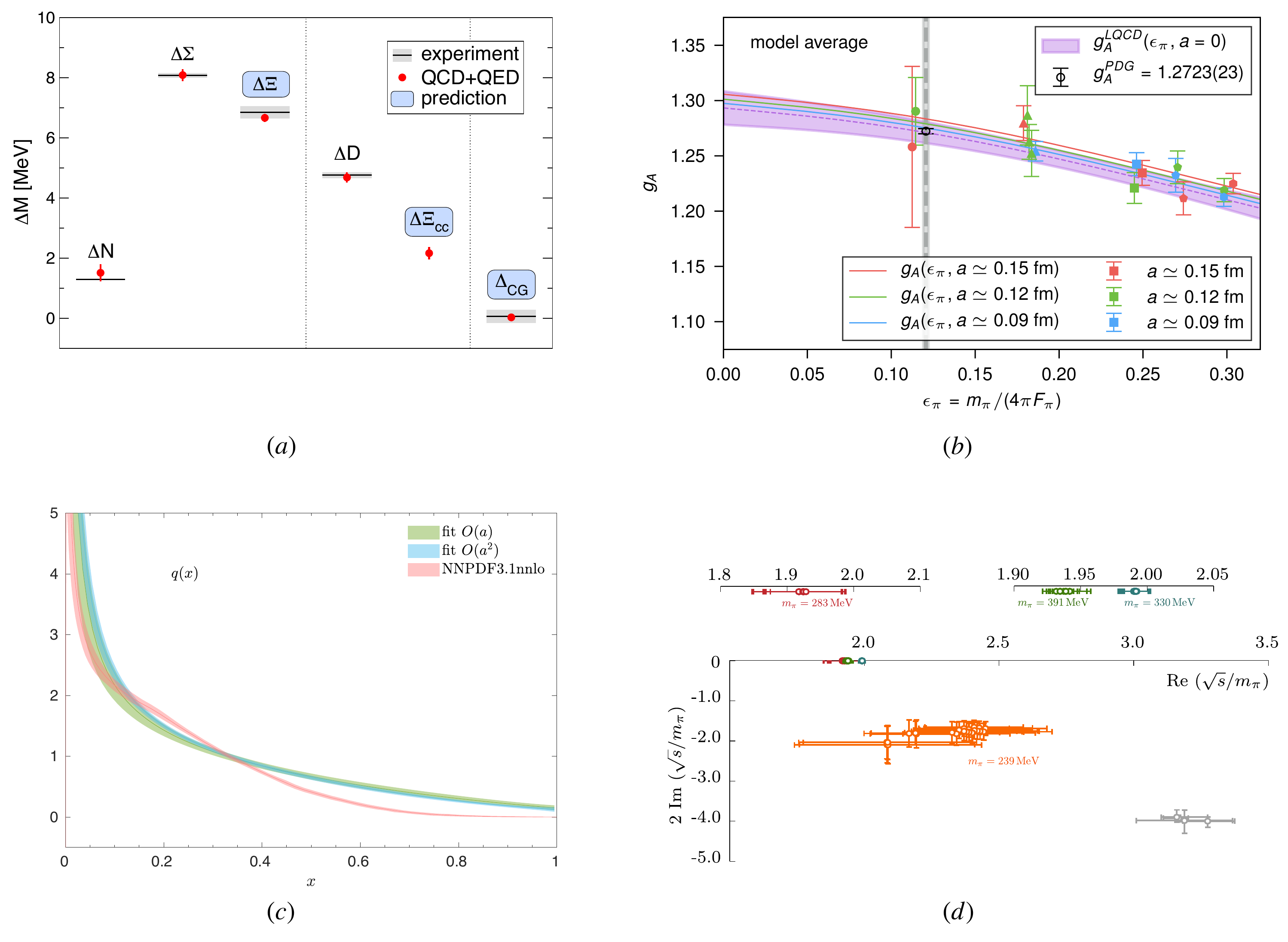}
    \caption{ Shown are some key highlights from the last decade. ($a$) Shown is the mass splitting between different isospin partners, including the neutron-proton, as calculated by Ref.~\cite{Borsanyi:2014jba}. ($b$) Shown is the first percent-level determination of the nucleon axial charge~\cite{Chang:2018uxx}. For a recent update of this calculation, see Ref.~\cite{Hall:2025ytt}. ($c$) Shown is a continuum-limit determination of the unpolarized isovector unpolarized PDF of the nucleon at a fixed value of the quark masses, corresponding to $m_\pi\approx 370$~MeV~\cite{Bhat:2022zrw}, compared to the phenomenological determination by the NNPDF Collaboration~\cite{NNPDF:2017mvq}.
    ($d$) Shown is the quark-mass evolution of the $\sigma$ resonance~\cite{Rodas:2023nec}, constrained from a series of calculations~\cite{Briceno:2016mjc,Briceno:2017qmb,Rodas:2023gma,Rodas:2023nec}. 
    }
    \label{fig:highlights}
\end{figure*}

The determination of masses of low-lying stable hadrons has reached a remarkable level of precision. A notable example is highlighted in Fig.~\ref{fig:highlights}($a$), where Ref.~\cite{Borsanyi:2014jba} calculated the mass splitting of a range of isospin partners, including the neutron-proton. This calculation included not only isospin-breaking effects due to the up- and down-quark mass splitting but also due to leading quantum electromagnetic (QED) effects. This introduces several complications, such as the infinite range of the electromagnetic interaction, the need for careful treatment of finite-volume effects, and the breaking of charge conservation by periodic boundary conditions. This is an ongoing area of research. 

In addition to these quantities, lattice QCD has achieved high-precision determinations of nucleon charges, including the axial charge ($g_A$), tensor charge ($g_T$), and scalar charge ($g_S$). These charges are essential for understanding processes such as neutron beta decay, searches for electric dipole moments, and probing possible physics beyond the Standard Model. A notable example is shown in Fig.~\ref{fig:highlights}($b$). There, one sees the percent-level calculation of $g_A$ using 2+1+1-flavor QCD at the physical pion mass by Ref.~\cite{Chang:2018uxx}. This calculation has been recently updated by the same collaboration in Ref.~\cite{Hall:2025ytt}. Other collaborations have provided precise results for $g_T$ and $g_S$~\cite{Bhattacharya:2016zcn,Gupta:2018qil}. These quantities are now computed with controlled systematics and play a vital role in nuclear and particle physics phenomenology.

\textbf{Hadron structure:}  
Beyond simple observables, attention has turned to more intricate properties of hadrons, particularly those that probe their internal structure. A central focus has been the calculation of electromagnetic form factors, which encode information about the spatial distributions of charge and magnetization within hadrons. Lattice QCD enables the determination of nucleon electric ($G_E$) and magnetic ($G_M$) form factors, as well as the axial form factor ($G_A$), which are directly related to experimentally measurable quantities like the charge radius, magnetic moment, and axial radius~\cite{Alexandrou:2017ypw,Green:2014xba,Capitani:2015sba}. These calculations have achieved increasing precision and are crucial for benchmarking QCD against experiment, providing input for neutrino-nucleus scattering models, and constraining searches for new physics.

More recently, attention has expanded to the calculation of parton distribution functions (PDFs) and related quantities, such as quasi-PDFs and pseudo-PDFs. These objects characterize the momentum and spin distributions of quarks and gluons within hadrons and are fundamental to understanding high-energy scattering processes. Traditionally, PDFs could only be accessed indirectly through global fits to experimental data, but novel approaches like large-momentum effective theory (LaMET)~\cite{Ji:2013dva,Radyushkin:2016hsy,Detmold:2005gg} have enabled their direct calculation on the lattice. These advances provide unprecedented insight into the internal structure of nucleons and other hadrons. A notable example of these studies is shown in Fig.~\ref{fig:highlights}($c$) by Ref.~\cite{Bhat:2022zrw}. There one sees the isovector unpolarized PDF of the nucleon obtained in the continuum limit for a set of unphysically heavy quark masses corresponding to $m_\pi\approx 370$~MeV. This PDF is a function of the momentum fraction ($x$) of the parton. Despite the unphysically heavy values of the quark masses, the results resemble the phenomenological extraction by the NNPDF Collaboration~\cite{NNPDF:2017mvq}. 

The field of hadron structure from lattice QCD is rapidly changing and growing. A final example of the breadth of this subfield is the gravitational form factors, which can provide insight into the distribution of energy, spin,
pressure, and shear forces within hadrons. A recent set of exploratory calculations~\cite{Hackett:2023rif,Hackett:2023nkr,Pefkou:2021fni} has demonstrated the viability of determining these via lattice QCD.

\textbf{Hadron spectroscopy:} One of the fields that substantially benefited from the advent of lattice QCD is hadron spectroscopy. The vast majority of states observed experimentally are hadronic resonances that decay into asymptotic states of two or more hadrons~\cite{ParticleDataGroup:2024}. These resonances can be rigorously defined as complex-valued poles in the scattering amplitudes of their byproducts. Because lattice QCD is defined in a finite Euclidean spacetime, scattering observables can not be obtained directly. Using correspondence between finite-volume observables and infinite-volume scattering observables, first identified by L\"uscher formalism~\cite{Luscher:1990ck}, a large number of such resonant states have and are being actively studied via lattice QCD, see Refs.~\cite{Dudek:2014qha,Whyte:2024ihh,Dudek:2012xn,Bulava:2016mks,BaryonScatteringBaSc:2023zvt,Padmanath:2022cvl} for some recent examples. 

In Fig.~\ref{fig:highlights}($d$), we see the notable example of the $\sigma$ resonance, the lightest resonance within QCD, which has been studied in Refs.~\cite{Briceno:2016mjc,Briceno:2017qmb,Rodas:2023gma,Rodas:2023nec}, among other investigations. The $\sigma$ decays approximately $100\%$ of the time to the $J^P=0^+$ $\pi\pi$ channel~\cite{ParticleDataGroup:2024}. Via the lattice, this state can be reconstructed by first obtaining the isoscalar $\pi\pi$ finite-volume spectrum. This is a highly non-trivial step since it requires the evaluation of so-called disconnected diagrams. This is made possible thanks to the advent of a novel smearing technique known as \emph{distilation}~\cite{HadronSpectrum:2009krc}.~\footnote{See Ref.~\cite{Lang:2024syy} for ongoing efforts aimed at further speeding up calculations performed using a distillation basis.} Once the spectrum has been obtained, one can then directly constrain the infinite-volume $\pi\pi$ amplitude for those same energies using the L\"uscher formalism and its extensions. The $\sigma$ then appears as a dynamically generated pole in this amplitude. If it is a bound state, it appears as a real-valued pole in the physical Riemann sheet of the amplitude. If it is a virtual bound state, it appears as a real-valued pole in the unphysical sheet. If it is a resonance, it appears as a complex-valued pole in the unphysical sheet. From Fig.~\ref{fig:highlights}($d$), one sees the evolution of this state from a real bound state, to a virtual bound state, to resonances as the quark masses approach the physical value. ~\footnote{See Refs.~\cite{Briceno:2017max,Hansen:2019nir,Mai:2021lwb} for recent reviews on the topic of hadron spectroscopy on the lattice.}

\textbf{Nuclear physics:}  Tools that have been developed for the study of the hadron spectrum are being adopted in the investigation of light nuclei on the lattice. Much attention has been paid to the deuteron and di-neutron channels. These have been primarily studied at unphysical heavy quark masses, where $m_\pi\sim 700$~MeV. Modern-day calculations suggest that at these unphysically heavy quark masses neither of these systems supports a bound state~\cite{BaSc:2025yhy,Amarasinghe:2021lqa,Aoki:2020bew}. Instead, they seem to have evidence of virtual bound states. Given the large statistical and systematic errors associated with these calculations, this is still a major line of research. 

These puzzling results have inspired, among other things, a novel line of research aimed at providing strict bounds on the spectra and matrix elements obtained from two- and three-point correlation functions~\cite{Wagman:2024rid,Hackett:2024xnx,Hackett:2024nbe} aimed at providing better constraints than the state of the art, namely the GEVP-based approach~\cite{Michael:1985ne,Luscher:1990ck}. This exciting line of research is still in its early stages.

\textbf{Finite temperature, chemical potential, and finite density:}  
The lattice QCD has achieved great success in studying QCD at nonzero temperature, providing first-principles insights into the properties of strongly interacting matter under extreme conditions. Notably, lattice calculations have constrained the QCD equation of state for zero chemical potential with high precision, mapping the transition from hadronic matter to the quark-gluon plasma and establishing that the transition at physical quark masses is a crossover rather than a true phase transition~\cite{Borsanyi:2013bia,HotQCD:2014kol}. 

Despite this progress, studying QCD at nonzero baryon chemical potential remains a major challenge due to the notorious sign problem: the fermion determinant becomes complex, precluding standard importance sampling techniques in Monte Carlo simulations. Various methods have been developed to circumvent or mitigate this problem, including reweighting, Taylor expansion in chemical potential, imaginary chemical potential approaches, and complex Langevin dynamics~\cite{deForcrand:2009zkb,Aarts:2015tyj}. While these techniques have enabled exploration of the QCD phase diagram at small chemical potentials, direct simulations at high baryon density remain out of reach. The search for a possible critical point in the QCD phase diagram is ongoing.

Recent years have seen significant advances in the exploration of finite-density QCD using novel algorithms and approaches. Notably, the application of Lefschetz thimble and generalized thimble methods~\cite{Cristoforetti:2012su, Alexandru:2016gsd} has opened new avenues for tackling the sign problem in toy models and, increasingly, in QCD-like theories. Furthermore, lattice calculations have begun to yield first-principles results for isosymmetric nuclear matter at moderate densities~\cite{Abbott:2024vhj,Abbott:2023coj}, marking a significant milestone towards ab initio studies of dense nuclear systems. These developments, while still in their early stages, demonstrate the vitality of the field and the promise of lattice methods for exploring the rich physics of QCD at finite density.

\textbf{Precision tests and the anomalous magnetic moment of the muon:}  
Lattice calculations play a key role in precision tests of the Standard Model by providing first-principles determinations of hadronic matrix elements that are essential for interpreting experimental measurements and searching for new physics. 

A prominent example is the determination of the hadronic contributions to the anomalous magnetic moment of the muon, $a_\mu = (g-2)_\mu/2$. The Standard Model prediction for $a_\mu$ relies on input from lattice QCD for the leading-order hadronic vacuum polarization (HVP) and hadronic light-by-light (HLbL) scattering contributions. Recent high-precision lattice calculations of the HVP~\cite{Borsanyi:2020mff,Chakraborty:2016mwy,Blum:2018mom,Ce:2022kxy,Alexandrou:2022amy, Boccaletti:2024guq} and HLbL~\cite{Blum:2019ugy,Gerardin:2019vio,Chao:2021tvp} terms are crucial for reducing the theoretical uncertainty and clarifying the longstanding discrepancy between experiment~\cite{Abi:2021gix, Bennett:2006fi} and the Standard Model prediction. The latest estimates of the HVP from the lattice drastically reduce the likelihood of possible BSM signals in $g-2$. 

In addition to $g-2$, lattice QCD is indispensable for precision studies of weak decays that test the flavor sector of the Standard Model and probe for new physics. Calculations of kaon decays, such as $K\rightarrow 2\pi$, provide direct input for understanding CP violation through the determination of the parameter $\epsilon'/\epsilon$~\cite{RBC:2015gro, RBC:2020kdj}. The RBC and UKQCD collaborations have achieved the first ab initio calculations of $K\rightarrow 2\pi$ amplitudes with controlled uncertainties, opening a new era for kaon physics.

Semileptonic decays of heavy mesons, such as $B\to D^{(*)}\ell\nu$ and $B\to\pi\ell\nu$, are also a major focus. Lattice QCD provides the form factors needed to extract CKM matrix elements, such as $|V_{cb}|$ and $|V_{ub}|$, from experimental measurements~\cite{Aoki:2024cqa, FermilabLattice:2014ysv,FermilabLattice:2015cdh,Meinel:2024pip,Flynn:2015mha,Na:2015kha}. Recently, exploratory calculations of multi-hadron weak decay are being performed~\cite{Leskovec:2025gsw}.   Progress in these calculations is essential for resolving current tensions in the unitarity of the CKM matrix and for searches for new physics in flavor observables.

Overall, continued advances in lattice QCD are vital for reducing theoretical uncertainties in these precision observables, enabling stringent tests of the Standard Model and providing sensitivity to possible new physics.

\textbf{Inclusive rates and real-time methods:}  
Traditional lattice simulations are formulated in Euclidean time, which enables the use of importance sampling and Monte Carlo techniques, but also fundamentally limits direct access to real-time (Minkowski) dynamics. As a result, key physical quantities such as transport coefficients, spectral functions, and real-time evolution of quantum systems are not immediately accessible from standard lattice QCD calculations.

To overcome these limitations, several innovative approaches have been developed. One would hope that by analytically continuing the Euclidean correlation functions computed on the lattice one could determine the real-time observables. However, the inversion of these relations is an ill-posed problem, highly sensitive to statistical noise and limited temporal extent of the lattice data. Techniques such as the Maximum Entropy Method~\cite{Asakawa:2000tr}, Bayesian inference~\cite{Burnier:2013nla}, and Backus-Gilbert reconstruction~\cite{Hansen:2019idp} are being explored to circumvent such limitations.

Another promising direction is the use of real-time lattice gauge theory simulations based on quantum computing and quantum simulation algorithms~
~\cite{Farrell:2024mgu, Turro:2024pxu, Jordan:2012xnu, Marshall:2015mna,Briceno:2020rar,  Davoudi:2024wyv,Jha:2024jan, DiMeglio:2023nsa,Bauer:2022hpo, Martinez:2016yna,Klco:2018kyo,Davoudi:2025rdv} using the Hamiltonian formulation of the theory~\cite{Kogut:1974ag}. Quantum computers, in principle, can efficiently simulate unitary time evolution of quantum field theories, potentially providing direct access to real-time phenomena such as scattering amplitudes, non-equilibrium dynamics, and quantum chaos. While current quantum hardware is still in its infancy, proof-of-principle demonstrations of digital quantum simulation of lattice gauge theories have already been realized in small systems.

Overall, the development of real-time and out-of-equilibrium methods is a vibrant interdisciplinary frontier, combining advances in numerical analysis, quantum information, and field theory. Continued progress in these directions promises to unlock new insights into dynamical phenomena in QCD and other quantum field theories.

\textbf{Algorithmic developments:}  
The last decade has seen remarkable progress in algorithms for lattice field theory, enabling simulations closer to the physical point, larger volumes, and finer lattice spacings. Among the most significant advances are the adoption of multigrid solvers~\cite{Babich:2010qb,Frommer:2013fsa}, which have dramatically reduced the cost of inverting the Dirac operator, particularly for light quark masses. Deflation techniques and hierarchical probing~\cite{Stathopoulos:2013aci,Frommer:2013fsa} have further improved the efficiency of solving large linear systems and estimating disconnected diagrams.

To address critical slowing down, especially near the continuum limit, new strategies have been developed, such as open boundary conditions~\cite{Luscher:2011kk}, master-field simulations~\cite{Luscher:2017cjh}, and parallel tempering~\cite{Hasenbusch:2017unr}. Multilevel Monte Carlo algorithms, originally developed for pure gauge theories, are now being generalized to full QCD~\cite{Ce:2016idq}.

In recent years, machine learning and artificial intelligence have emerged as promising tools in the field. Machine learning techniques are being explored to accelerate the generation of gauge configurations via generative models such as normalizing flows~\cite{Albergo:2019eim,Boyda:2022nmh,Abbott:2022zsh}, to optimize sampling and reduce autocorrelations, and to reconstruct spectral functions from Euclidean correlators~\cite{Shanahan:2018vcv}. Neural networks are also being explored for parameter inference, noise reduction, and feature extraction from large lattice datasets.

These algorithmic innovations, in combination with advances in hardware and software frameworks, are pushing the boundaries of what is possible in lattice field theory, enabling new physics studies and higher-precision results than ever before.

\textbf{Lattice field theory beyond QCD: }
While lattice gauge theory was originally developed to study QCD, its methods broadly apply to a wide range of quantum field theories, including those that are candidates for physics beyond the Standard Model (BSM). 
One prominent example is the study of composite Higgs models and technicolor theories, where the Higgs boson arises as a bound state of new strong dynamics. Lattice calculations have been used to explore the phase structure, spectrum, and conformal properties of SU($N$) gauge theories with various numbers of fermion flavors and representations~\cite{DeGrand:2015zxa,Appelquist:2016viq}. These studies help constrain the viability of candidate BSM theories and provide input for model building.
Lattice methods have also been employed to investigate dark matter scenarios involving strongly-coupled hidden sector models~\cite{Kribs:2016cew}. By computing the spectrum and interactions of the hypothetical dark sector, lattice field theory can provide predictions for relic abundances, self-interactions, and possible signals in direct or indirect detection experiments.
Furthermore, lattice techniques are being applied to supersymmetric gauge theories, scalar field theories, and models with topological terms, broadening the scope of non-perturbative studies in quantum field theory. As computational resources and algorithms continue to advance, lattice field theory will remain a powerful approach for exploring the rich landscape of QFTs, including those with potential implications for new physics.

\vspace{0.5em}
\textbf{Summary: } In summary, modern-day research in lattice field theory spans a vast landscape, from precision calculations of simple observables to the exploration of complex systems and the development of novel algorithms. The field continues to evolve rapidly, driven by both theoretical innovation and advances in computational technology.

\begin{ack}[Acknowledgments]%
RAB was partly supported by the U.S. Department of Energy, Office of Science, Office of Nuclear Physics under Award No. DE-SC0025665 and No. DE-AC02-05CH11231. RAB would like to thank D. Pefkou, F. Ortega-Gama, and I. Burbano for feedback on the manuscript. 
\end{ack}


\bibliographystyle{elsarticle-num}
\bibliography{bibi.bib}

\end{document}